\documentclass[twocolumn,superscriptaddress,floatfix,preprintnumbers]{revtex4}
\usepackage{graphics,amssymb,amsmath,color}
\usepackage{graphicx}

\DeclareMathOperator{\diag}{diag}
\DeclareMathOperator{\minimize}{minimize}
\DeclareMathOperator{\subjectto}{subject\;to}

\begin{document}

\title{Fabrication-constrained nanophotonic inverse design}
\author{Alexander Y. Piggott}
\author{Jan Petykiewicz}
\author{Logan Su}
\author{Jelena Vu\v{c}kovi\'{c}}
\affiliation{Ginzton Laboratory, Stanford University, Stanford, California 94305, USA}

\date{\today}
\begin{abstract}
A major difficulty in applying computational design methods to nanophotonic devices is ensuring that the resulting designs are fabricable. Here, we describe a general inverse design algorithm for nanophotonic devices that directly incorporates fabrication constraints. To demonstrate the capabilities of our method, we designed a spatial-mode demultiplexer, wavelength demultiplexer, and directional coupler. We also designed and experimentally demonstrated a compact, broadband $1 \times 3$ power splitter on a silicon photonics platform. The splitter has a footprint of only $3.8 \times 2.5~\mathrm{\mu m}$, and is well within the design rules of a typical silicon photonics process, with a minimum radius of curvature of $100~\mathrm{nm}$. Averaged over the designed wavelength range of $1400 - 1700~\mathrm{nm}$, our splitter has a measured insertion loss of  $0.642 \pm 0.057 ~\mathrm{dB}$ and power uniformity of $0.641 \pm 0.054~\mathrm{dB}$.
\end{abstract}

\maketitle

\section{Introduction}
Nanophotonic devices are typically designed by starting with an analytically designed structure, and hand-tuning a few parameters \cite{gtreed_2008}. In recent years, it has become increasingly popular to automate this process with the use of powerful optimization algorithms. In particular, by searching the full space of possible structures, it is possible to design devices with higher performance and smaller footprints than traditional devices \cite{amutapcica_eo2009, jjensen_lpr2011, lalau-keraly_oe2013, jlu_oe2013, aniederberger_oe2014, aypiggott_np2015, lffrellsen_oe2016}.

A major challenge when designing devices with arbitrary topologies is ensuring that the structures remain fabricable. Many of these computationally designed structures have excellent performance when fabricated using high-resolution electron-beam lithography, but they have features which are difficult to resolve with industry-standard optical lithography \cite{jjensen_lpr2011, aypiggott_np2015, lffrellsen_oe2016}.

Building on our previous work \cite{jlu_oe2013, aypiggott_sr2014, aypiggott_np2015}, we propose an inverse design method for nanophotonic devices that incorporates fabrication constraints. Our algorithm achieves an approximate minimum feature size by imposing curvature constraints on dielectric boundaries in the structure. We then demonstrate the capabilities of our method by designing a spatial-mode demultiplexer, wavelength demultiplexer, and directional coupler, and experimentally demonstrating an ultra-broadband $1 \times 3$ power splitter. All of our designs are compact, have no small features, and should be resolvable using modern photolithography. Additionally, with the exception of the wavelength demultiplexer, all of our devices are well within the design rules of existing silicon photonics processes.

\section{Design Method}

Due to the complexity of accurately modelling lithography and etching processes, most attempts to incorporate fabrication constraints into computational nanophotonic design have focused on heuristic methods. One approach is to restrict the design to rectangular pixels which are larger than the mininum allowable feature size \cite{bshen_np2015}. The resulting Manhattan geometry, however, is restrictive and likely not optimal for optical devices. Another method involves applying a convolutional filter to the design followed by thresholding \cite{yelesin_pnfa2012, ydeng_prsa2016, lhfrandsen_spie2016}, which can introduce artifacts smaller than the desired feature size. The approach used in this work is to impose curvature constraints on the device boundaries, which avoids the aforementioned issues. Curvature limits have been successfully applied in earlier work \cite{lalau-keraly_oe2013}, but were not described in detail nor validated with experimental demonstrations.

\subsection{Level Set Formulation}
We assume that our device is planar and consists of only two materials. We can represent our structure by constructing a continuous function $\phi(x,y):\mathbb{R}^2 \rightarrow \mathbb{R}$ over our design region, and letting the boundaries between the materials lie on the level set $\phi = 0$. The permittivity $\epsilon$ is then given by
\begin{align}
\epsilon(x,y) = 
\begin{cases}
  \epsilon_1 & \mathrm{for} \; \phi(x,y) \leq 0 \\
  \epsilon_2 & \mathrm{for} \; \phi(x,y) > 0. \\
\end{cases}
\end{align}
The advantage of this implicit representation is that changes in topology, such as the merging and splitting of holes, are trivial to handle. We can also manipulate our structure by adding a time dependence, and evolving $\phi(x,y,t)$ as a function of time $t$ with a variety of partial differential equations collectively known as level set methods \cite{sosher_2003, mburger_ejam2005}.

To design a device, we first choose some objective function $f[\epsilon]$ which describes how well the structure matches our electromagnetic performance constraints \cite{jlu_oe2013,aypiggott_np2015}. We then evolve our structure, represented by $\phi$, in such a way that we minimize our objective $f$. We can achieve this by adapting gradient descent optimization to our level set representation. The level set equation for moving boundaries in the normal direction is
\begin{align}
\phi_t + v(x,y) \left| \nabla \phi \right| = 0
\label{eqn:wgSplit3_normprop}
\end{align}
where $\nabla \phi = \phi_x + \phi_y$ is the spatial gradient of $\phi$, and $v(x,y)$ is the local velocity. To implement gradient descent, we choose the velocity field $v(x,y)$ to correspond to the gradient of the objective function $f[\epsilon]$ \cite{mburger_ejam2005}. The gradient can be efficiently computed using adjoint sensitivity analysis \cite{jjensen_lpr2011, lalau-keraly_oe2013, aypiggott_sr2014, aniederberger_oe2014}. As $t \rightarrow \infty$, $\phi$ converges to a locally optimal structure.

Unfortunately, this approach tends to result in the formation of extremely small features. We can avoid this problem by periodically enforcing curvature constraints. The level set equation for smoothing out curved regions is
\begin{align}
\phi_t - \kappa \left| \nabla \phi \right| = 0
\label{eqn:wgSplit3_meancurv}
\end{align}
where the local curvature $\kappa$ is given by
\begin{align}
\kappa = \nabla \cdot \left(\frac{\nabla \phi}{\left| \nabla \phi \right|} \right) = \frac{\phi_x^2 \phi_{yy} - 2 \phi_x \phi_y \phi_{xy} + \phi_{xx} \phi_y^2}{\left|\nabla \phi \right|^3}.
\end{align}
Although equation \ref{eqn:wgSplit3_meancurv} removes highly curved regions more quickly \cite{sosher_2003}, the boundaries are eventually reduced to a set of straight lines with zero curvature as $t \rightarrow \infty$.

From a fabrication perspective, we only need to smooth regions which are above some maximum allowable curvature $\kappa_0$. We can do this by introducing a weighting function
\begin{align}
b(\kappa) = 
\begin{cases}
  1 & \mathrm{for} \; |\kappa| > \kappa_0\\
  0 & \mathrm{otherwise} \\
\end{cases}
\end{align}
and modifying equation \ref{eqn:wgSplit3_meancurv} to be
\begin{align}
\phi_t - b(\kappa) \kappa \left| \nabla \phi \right| = 0.
\label{eqn:wgSplit3_curvfilt}
\end{align}
If we evolve $\phi$ with equation \ref{eqn:wgSplit3_curvfilt} until it reaches steady state, the maximum curvature will be less than or equal to $\kappa_0$.

Although curvature limiting will eliminate the formation of most small features, it does not prevent the formation of narrow gaps or bridges. We detect these features by applying morphological dilation and erosion operations to the set $\phi > 0$, and checking for changes in topology. Once detected, these narrow gaps and bridges can be eliminated by ``cutting'' them in half, and then applying curvature filtering to round out the sharp edges.

The final design algorithm is as follows:

\begin{enumerate}
\item Initialize $\phi$ and $\delta t$.
\item Repeat until $\delta t < \delta t_{min}$.
\begin{enumerate}
  \item Let $\phi' \leftarrow \phi$.
  \item \textbf{Gradient descent}: evolve $\phi'$ with eqn. \ref{eqn:wgSplit3_normprop} for time $\delta t$.
	\item \textbf{Gap and bridge removal}: detect any small gaps or bridges, and modify $\phi'$ to remove them.
  \item \textbf{Curvature limit}: evolve $\phi'$ with eqn. \ref{eqn:wgSplit3_curvfilt} until convergence.
  \item \textbf{If} $f\left[\epsilon[\phi']\right] < f\left[\epsilon[\phi]\right]$, \textbf{then} let $\phi \leftarrow \phi'$ and increase $\delta t$. \\
\textbf{Otherwise}, decrease $\delta t$.
\end{enumerate}
\end{enumerate}

A detailed description of the objective function $f[\epsilon]$ and implementation details can be found in the supplementary information.

\section{Designed Devices}
To demonstrate the capabilities of our design method, we designed a variety of three-dimensional waveguide-coupled devices on a silicon photonics platform. All of the structures we show here consist of a single fully-etched $220~\mathrm{nm}$ thick $\mathrm{Si}$ layer with $\mathrm{SiO}_2$ cladding. Refractive indices of $n_\mathrm{Si} = 3.48$ and $n_\mathrm{SiO_2} = 1.44$ were used.

\subsection{$1 \times 3$ splitter}
Our first device is a broadband $1 \times 3$ power splitter with $500~\mathrm{nm}$ wide input and output waveguides. We constrained the mininum radius of curvature to be $100~\mathrm{nm}$, well within the typical design rules of a silicon photonics process, and enforced bilateral symmetry. To design the splitter, we specified that power in the fundamental traverse-electric (TE) mode of the input waveguide should be equally split into the fundamental TE mode of the three output waveguides, with at least $95\%$ efficiency. Broadband performance was achieved by simultaneously optimizing at 6 equally spaced wavelengths from $1400 - 1700~\mathrm{nm}$. 

\begin{figure*}[htbp]
\centering
\includegraphics[width=0.85\textwidth]{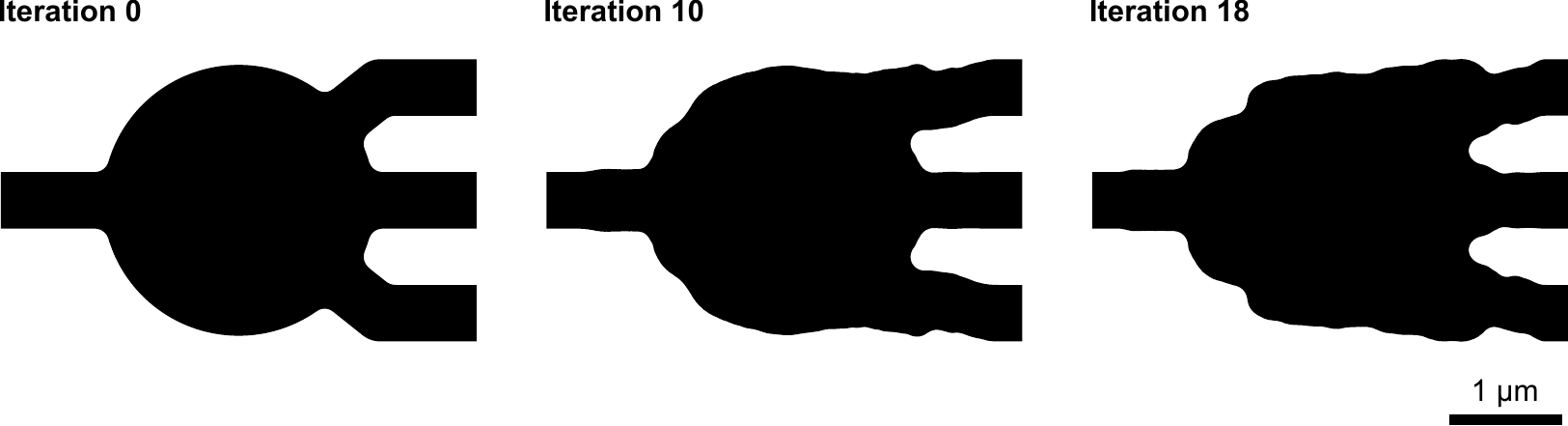}
\caption{Intermediate steps in the optimization process for the $1 \times 3$ splitter. Starting with a star-shaped geometry, the optimization converged in 18 iterations. The minimum radius of curvature in the design was set to $100~\mathrm{nm}$. Regions with Si are denoted in black, and $\mathrm{SiO}_2$ is denoted in white.}
\label{fig:1_wgSplit3_opt}
\vspace{-0.5cm}
\end{figure*}

The optimization process is illustrated in figure \ref{fig:1_wgSplit3_opt}, and the simulated electromagnetic fields and simulated performance are shown in figures \ref{fig:2_wgSplit3_dev} and \ref{fig:3_wgSplit3_S-params}. Starting with a star-shaped geometry, the optimization process converged in 18 iterations. Each iteration required two electromagnetic simulations per design frequency (see supplementary information), resulting in a total of 216 simulations. The device was designed in approximately 2 hours on a single server with an Intel Core i7-5820K processor, 64GB of RAM, and three Nvidia Titan Z graphics cards. Since the computational cost of optimization is dominated by the electromagnetic simulations, we performed them using a graphical processing unit (GPU) accelerated implementation of the finite-difference frequency-domain (FDFD) method \cite{wshin_jcp2012, wshin_oe2013}, with a spatial step size of $40~\mathrm{nm}$. A single FDFD solve is considerably faster and less computationally expensive than a finite-difference time-domain (FDTD) simulation.

Interestingly, the splitter appears to be operate using the multi-mode interferometer (MMI) principle \cite{pabesse_jlt1994}, with a geometry that resembles a boundary-optimized MMI. This was without any human input or intervention throughout the design process, suggesting that MMI-based devices may be optimal for this particular application. 

\subsection{Spatial mode demultiplexer}
Our second device is a spatial-mode demultiplexer that takes the $\mathrm{TE}_{10}$ and $\mathrm{TE}_{20}$ modes of a $750~\mathrm{nm}$ wide input waveguide, and routes them to the fundamental TE mode of two $400~\mathrm{nm}$ wide output waveguides. To design this device, we specified that $> 90\%$ of the input power should be transmitted to the desired output port, and $< 1\%$ should be coupled into the other output. As with the $1 \times 3$ splitter, this device was designed to be broadband by optimizing at six evenly spaced wavelengths between $1400~\mathrm{nm}$ and $1700~\mathrm{nm}$.  To obtain an initial structure for the level set optimization, we started with a uniform permittivity in the design region, allowed the permittivity to vary continuously in the initial stage of optimization, and applied thresholding to obtain a binary structure \cite{jlu_oe2013}. We used a minimum radius of curvature of $70~\mathrm{nm}$, and a minimum gap or bridge width of $90~\mathrm{nm}$. 

\begin{figure*}[htbp]
\centering
\includegraphics[scale=0.57]{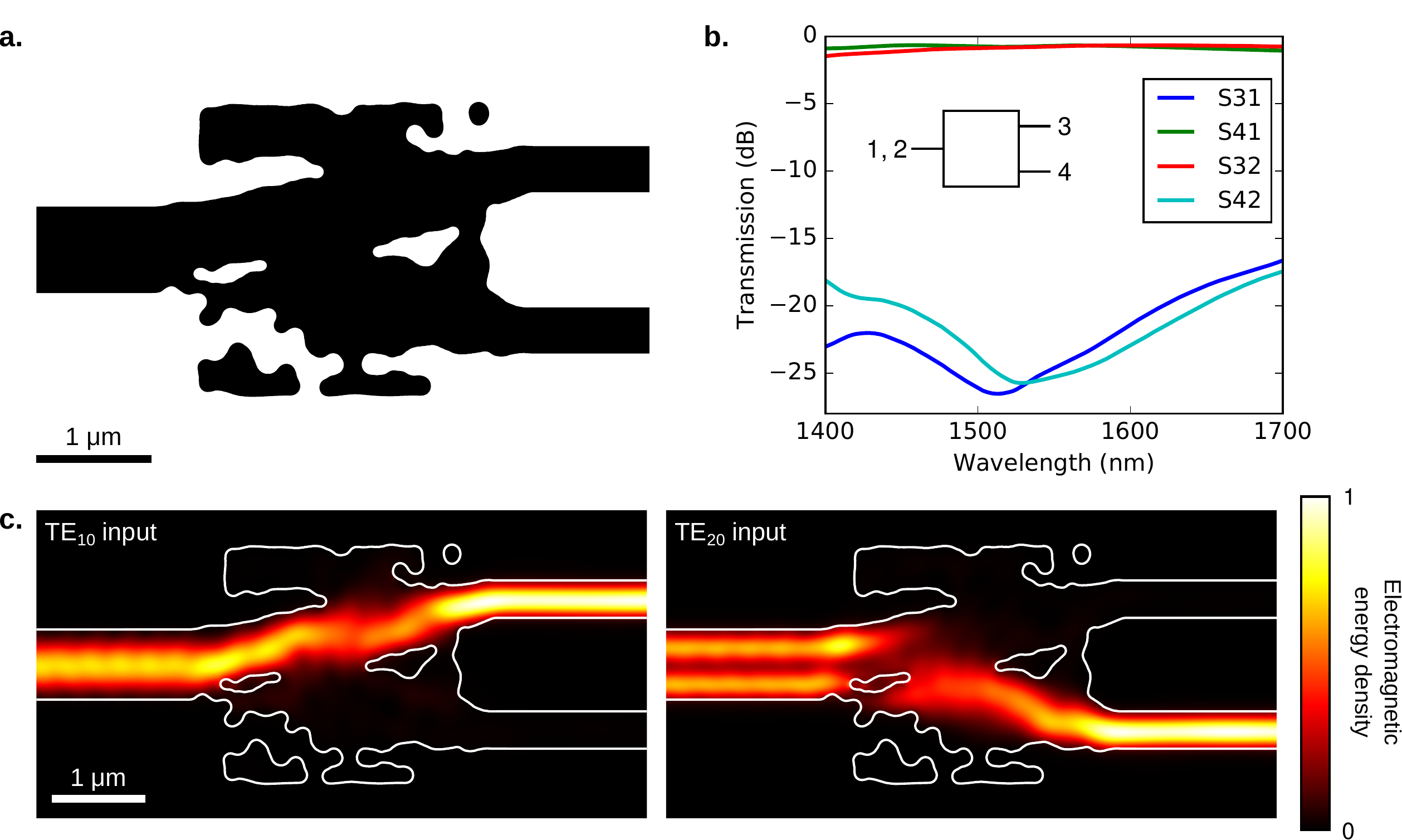}
\caption{A spatial mode demultiplexer that takes the $\mathrm{TE}_{10}$ and $\mathrm{TE}_{20}$ modes of a $750~\mathrm{nm}$ wide input waveguide, and routes them to the $\mathrm{TE}_{10}$ mode of two $400~\mathrm{nm}$ wide output waveguides. Here, we present \textbf{(a)} the final design, \textbf{(b)} simulated S-parameters, and \textbf{(c)} the electromagnetic energy density $U = \frac{1}{2} \epsilon E^2 + \frac{1}{2} \mu H^2$ at $1550~\mathrm{nm}$, where $\epsilon$ and $\mu$ are the permittivity and permeability, and $E$ and $H$ are the electric and magnetic fields respectively. The fields and S-parameters were calculated using finite-difference time-domain (FDTD) simulations. The boundaries of the device are outlined in white.}
\label{fig:wgSDM}
\end{figure*}

The final design and simulated performance are illustrated in figure \ref{fig:wgSDM}. The spatial mode multiplexer has an average insertion loss of $0.826~\mathrm{dB}$, and a contrast better than $16~\mathrm{dB}$ over the design bandwidth of $1400 - 1700~\mathrm{nm}$. 


\subsection{Wavelength demultiplexer}
Our third device is a 3-channel wavelength demultiplexer with a $40~\mathrm{nm}$ channel spacing with $500~\mathrm{nm}$ wide input and output waveguides. To design this device, we specified that $> 80\%$ of the input power should be transmitted to the desired output port, and $< 1\%$ should be coupled into the remaining outputs. The initial structure was a rectangular slab of silicon with a regular array of holes, which had a pitch of $400~\mathrm{nm}$ and a diameter of $250~\mathrm{nm}$. We enforced a minimum radius of curvature of $40~\mathrm{nm}$, and a minimum gap or bridge width of $90~\mathrm{nm}$. 

\begin{figure*}[htbp]
\centering
\includegraphics[scale=0.57]{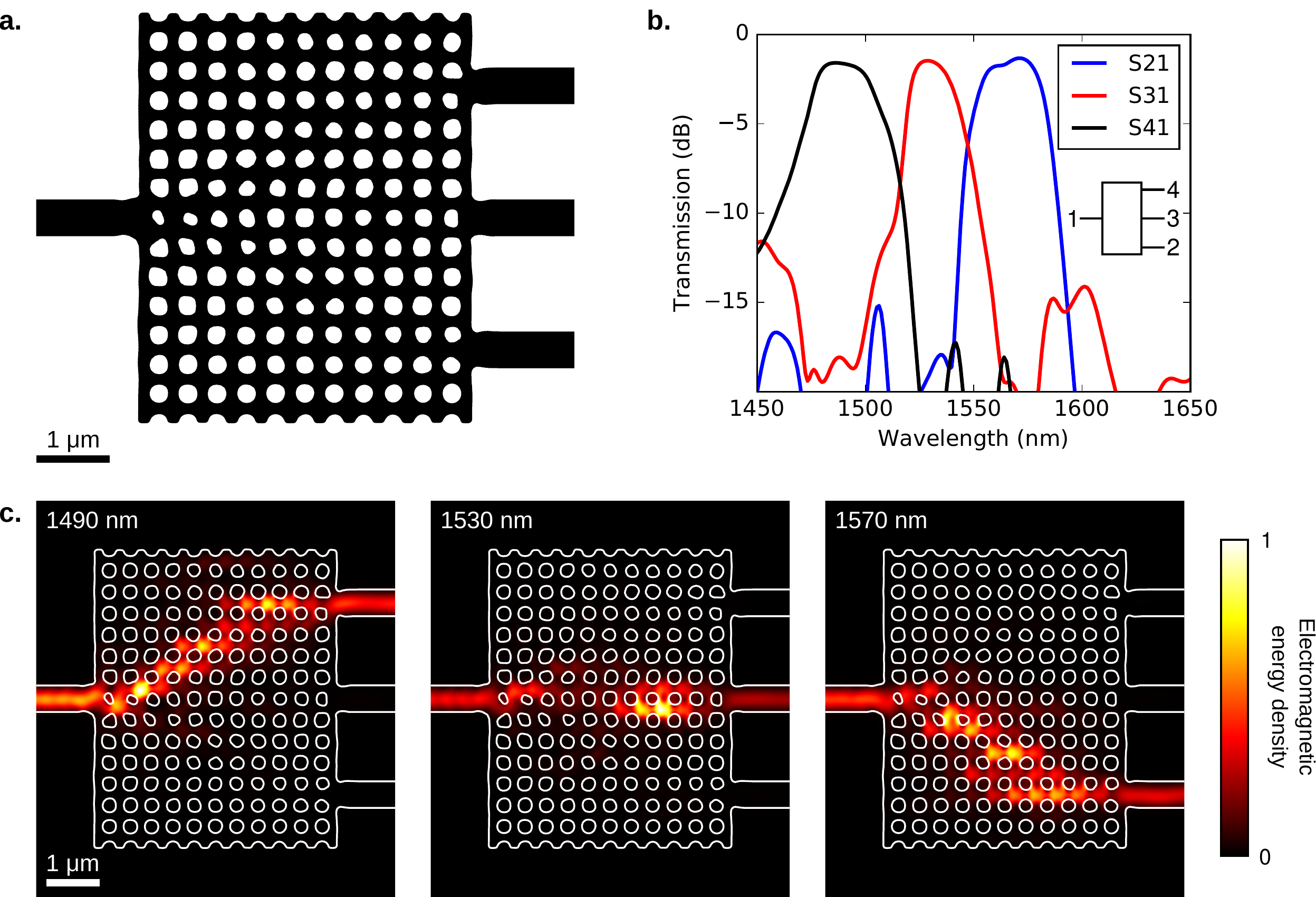}
\caption{A compact 3-channel wavelength demultiplexer with a $40~\mathrm{nm}$ channel spacing. The input and output waveguides are all $500~\mathrm{nm}$ wide. Here, we present \textbf{(a)} the final design, \textbf{(b)} simulated S-parameters, and \textbf{(c)} the electromagnetic energy density at the three operating wavelengths. }
\label{fig:wgWDM3}
\end{figure*}

The final design and simulated performance are illustrated in figure \ref{fig:wgWDM3}. At the center of each channel, the insertion loss is approximately $1.5~\mathrm{dB}$, and the contrast is better than $16~\mathrm{dB}$. Each channel has a usable bandwidth $> 10~\mathrm{nm}$.


\subsection{Directional coupler}
Our final device is a relatively compact 50-50 directional coupler, with $400~\mathrm{nm}$ input and output waveguides. This device was designed by specifying that half the power in fundamental mode of the input waveguide should be coupled into each of the outputs, with $> 90\%$ efficiency. As in the design of the spatial mode demultiplexer, we obtained an initial structure by starting with a uniform permittivity in the design region, allowing the permittivity to vary continuously in the initial stage of optimization, and applying thresholding. To achieve moderate broadband performance, the device was simultaneously optimized for 6 wavelengths between $1470 - 1630~\mathrm{nm}$. We enforced a minimum radius of curvature of $70~\mathrm{nm}$, and a minimum bridge width of $90~\mathrm{nm}$. 

\begin{figure*}[htbp]
\centering
\includegraphics[scale=0.57]{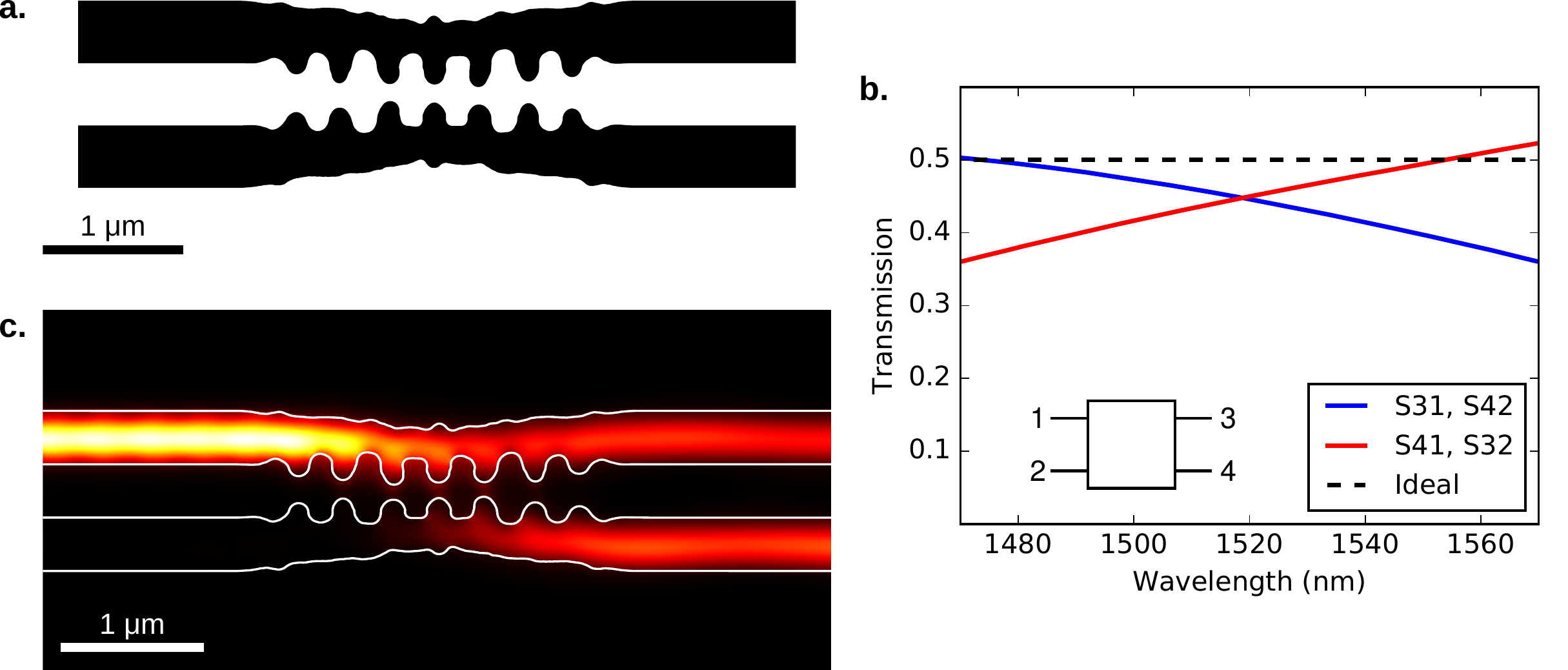}
\caption{A relatively compact and broadband $50-50$ directional coupler with $400~\mathrm{nm}$ input and output waveguides. The device resembles a grating-assisted directional coupler. Here, we present \textbf{(a)} the final design, \textbf{(b)} simulated S-parameters, and \textbf{(c)} the electromagnetic energy density at $1550~\mathrm{nm}$. }
\label{fig:wg90deghybrid}
\end{figure*}

The final device and simulated performance are illustrated in figure \ref{fig:wg90deghybrid}. At the optimal operating point of $1520~\mathrm{nm}$, the device couples $90\%$ of the input power into the desired output waveguides. The device structure appears to be a grating-assisted directional coupler.


\section{Experimental Realization of $1 \times 3$ Splitter}
Robust and efficient power splitters are essential building blocks for integrated photonics. A variety of $1 \times 2$ splitters with attractive performance have been demonstrated on the silicon photonics platform, ranging from conventional devices \cite{asakai_ieicete2002, shtao_oe2008} to those designed using advanced optimization techniques \cite{piborel_el2005,lalau-keraly_oe2013,yzhang_oe2013}. However, it is not possible to split power equally into an arbitrary number of waveguides by cascading $1 \times 2$ splitters, and efficient and compact devices that fill this gap are lacking in the literature. To help fill this gap, we fabricated and experimentally demonstrated the $1 \times 3$ splitter we presented in the previous section. Our $1 \times 3$ splitter is considerably smaller and more broadband than any existing device in the literature \cite{pabesse_jlt1996,mzhang_oe2010}.

\begin{figure*}[htbp]
\centering
\includegraphics[width=\textwidth]{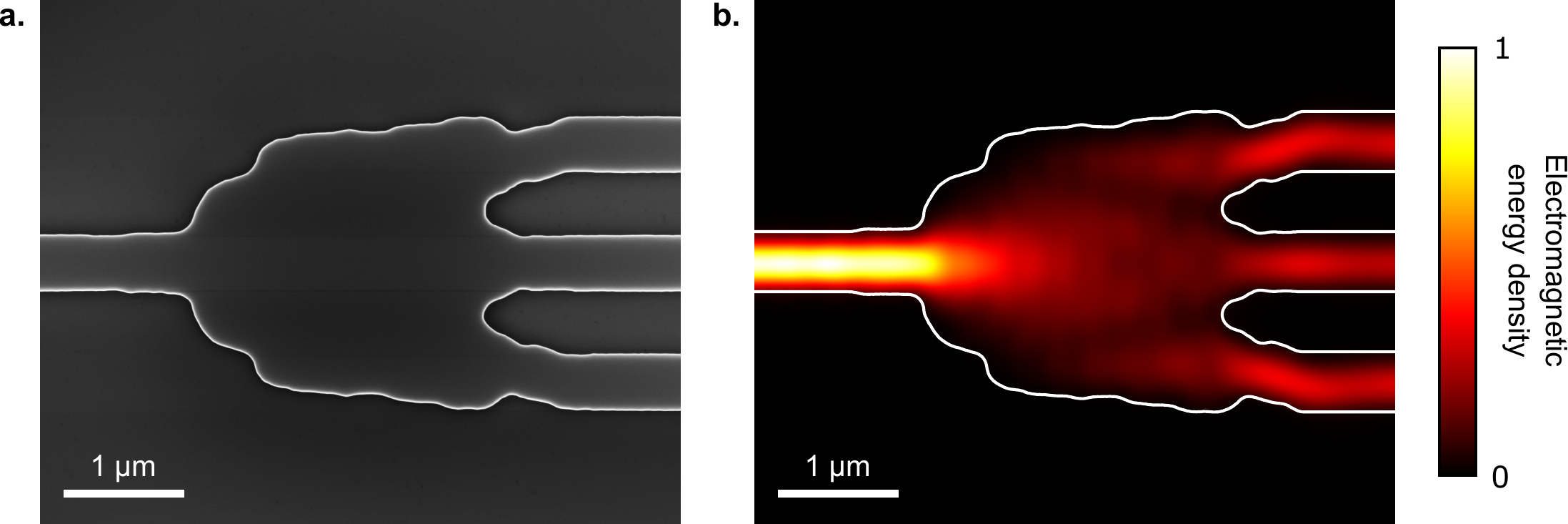}
\caption{The broadband $1 \times 3$ splitter. \textbf{(a)} SEM image of the fabricated splitter. The device was made by fully etching the $220~\mathrm{nm}$ device layer of an SOI wafer. The total footprint is $3.8 \times 2.5~\mathrm{\mu m}$. This image was taken before the devices were capped with oxide. \textbf{(b)} Electromagnetic energy density in the device at $1550~\mathrm{nm}$.}
\label{fig:2_wgSplit3_dev}
\end{figure*}

\subsection{Fabrication}
The power splitters were fabricated on Unibond SmartCut silicon-on-insulator (SOI) wafers obtained from SOITEC, with a nominal $220~\mathrm{nm}$ device layer, and $3.0~\mathrm{\mu m}$ buried oxide layer. A JEOL JBX-6300FS electron-beam lithography system was used to pattern a $330~\mathrm{nm}$ thick layer of ZEP-520A resist spun on the samples. A transformer-coupled plasma etcher was used to transfer the pattern to the device layer, using a $\mathrm{C}_2 \mathrm{F}_6$ breakthrough step and $\mathrm{BCl}_3 / \mathrm{Cl}_2 / \mathrm{O}_2$ main etch. The mask was stripped by soaking in solvents, followed by a piranha $(\mathrm{H}_2\mathrm{SO}_4 / \mathrm{H}_2\mathrm{O}_2)$ clean. Finally, the devices were capped with $1.6~\mathrm{\mu m}$ of LPCVD (low pressure chemical vapour deposition) oxide.

A multi-step etch-based process was used to expose waveguide facets for edge coupling. First, a chrome mask was deposited using liftoff to protect the devices. Next, the oxide cladding, device layer, and buried oxide layer were etched in a inductively-coupled plasma etcher using a $\mathrm{C}_4 \mathrm{F}_8 / \mathrm{Ar} \mathrm{O}_2$ chemistry. To provide mechanical clearance for the optical fibers, the silicon substrate was then etched to a depth of $\sim 100~\mathrm{\mu m}$ using the Bosch process in a deep reactive-ion etcher (DRIE). Finally, the chrome mask was chemically stripped, and the samples were diced into conveniently-sized pieces.

\subsection{Characterization}
The final splitter is illustrated in figure \ref{fig:2_wgSplit3_dev}, showing both an scanning-electron micrograph (SEM) of the fabricated device, and simulated electromagnetic energy density at the center wavelength of $1550~\mathrm{nm}$.

Transmission through the device was measured by edge-coupling to the input and output waveguides using lensed fibers. A polarization-maintaining fiber was used on the input side to ensure that only the TE mode of the waveguide was excited. To obtain consistent coupling regardless of the transmission spectra of the devices, the fibers were aligned by optimizing the transmitted power of a $1570~\mathrm{nm}$ laser. The transmission spectrum was then measured by using a supercontinuum source and a spectrum analyzer. The device characteristics were obtained by normalizing the transmission with respect to a waveguide running parallel to the device.

\begin{figure*}[htbp]
\centering
\includegraphics[width=\textwidth]{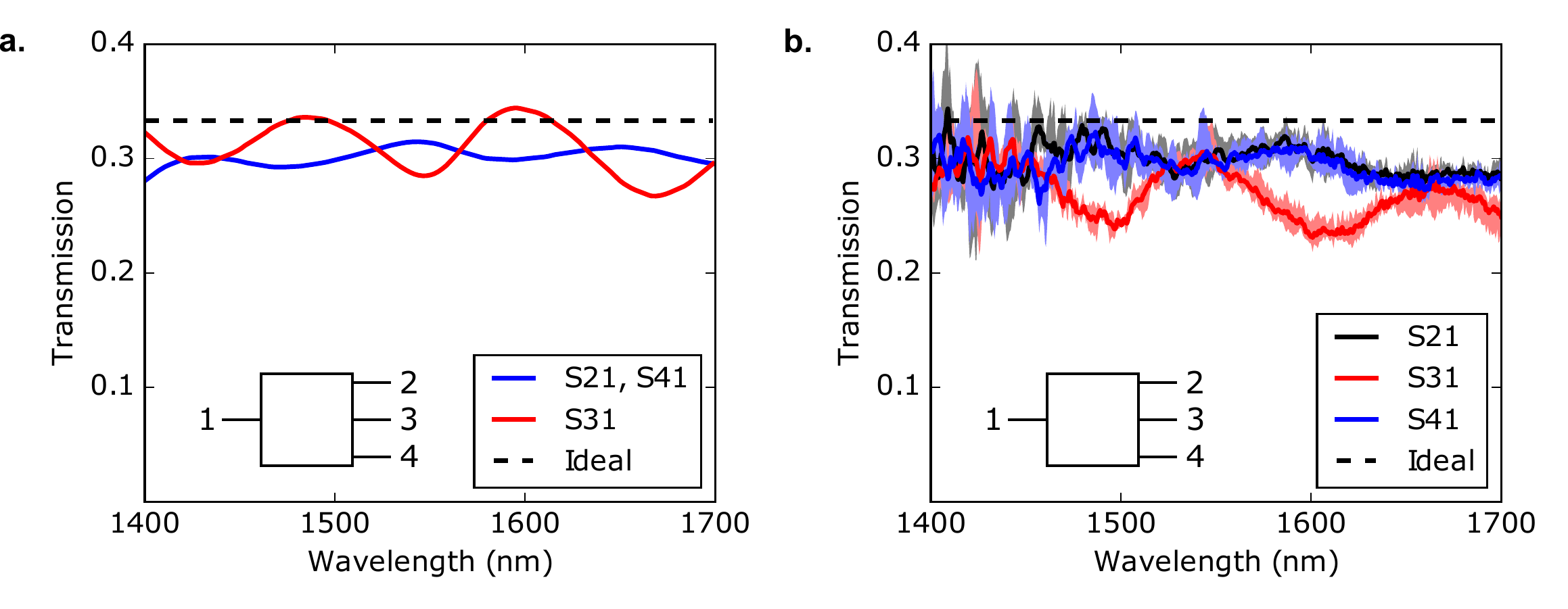}
\caption{Simulated and measured S-parameters for the broadband $1 \times 3$ splitter, where $Sij$ is the transmission from port $i$ to port $j$. \textbf{(a)} Simulated performance, calculated using finite-difference time-domain (FDTD) simulations. Due to bilateral symmetry in the structure, S21 and S41 are equal to each other. \textbf{(b)} Measured device performance. Here, we have overlaid the measurements for 4 identically fabricated devices. The average values are denoted by the solid lines, and the minimum and maximum values are denoted by the shaded areas.}
\label{fig:3_wgSplit3_S-params}
\end{figure*}

The simulated and measured transmission spectra of the device are plotted in figure \ref{fig:3_wgSplit3_S-params}. The simulations and measurements match reasonably well, although the measured devices have slightly higher losses and exhibit a spectral shift with respect to simulations. The device performance is highly consistent across all 4 measured devices, indicating that they are robust to fabrication error. The spectral shifts are likely due to slight over-etching or under-etching errors, as indicated by simulations we present in the supplementary information. 

The two key criteria for a power splitter are low insertion loss, and excellent power uniformity. The power uniformity is defined as the ratio between the maximum and minimum output powers. Averaged over the designed wavelength range of $1400 - 1700~\mathrm{nm}$, our $1 \times 3$ splitter has a measured insertion loss of $0.642 \pm 0.057 ~\mathrm{dB}$, and a power uniformity of $0.641 \pm 0.054~\mathrm{dB}$. Here, the uncertainty refers to the variability between different measured devices.

\section{Conclusion}
In summary, we have incorporated fabrication constraints into an inverse design algorithm for nanophotonic devices. Using this method, we designed a spatial mode demultiplexer, a 3-channel wavelength demultiplexer, and 50-50 directional coupler. We also designed and experimentally demonstrated a broadband $1 \times 3$ splitter. Critically, our devices have no small features which would be difficult to resolve with photolithography, paving the way for inverse designed structures to become practical components of integrated photonics systems.

\section*{Acknowledgments}
This work was funded by the AFOSR MURI for Aperiodic Silicon Photonics, grant number FA9550-15-1-0335, the Gordon and Betty Moore Foundation, and GlobalFoundries Inc. All devices were fabricated at the Stanford Nanofabrication Facility (SNF) and Stanford Nano Shared Facilities (SNSF).

\section*{Author contributions statement}
A.Y.P. designed, simulated, fabricated and measured the devices. A.Y.P., J.P., and L.S. developed the design and simulation software. J.V. supervised the project. All members contributed to the discussion and analysis of the results.

\section*{Additional information}
The authors declare no competing financial interests.

\vfill


\clearpage
\onecolumngrid

\begin{center}
\textbf{\large Supplemental Information}
\end{center}

\twocolumngrid
\setcounter{equation}{0}
\setcounter{figure}{0}
\setcounter{table}{0}
\setcounter{page}{1}
\makeatletter
\renewcommand{\theequation}{S\arabic{equation}}
\renewcommand{\thefigure}{S\arabic{figure}}
\renewcommand{\bibnumfmt}[1]{[S#1]}
\renewcommand{\citenumfont}[1]{S#1}

\section{Electromagnetic Design Method}
\subsection{Problem description}
Our goal is to automate the design of all passive photonic structures. Thus, our first task is to come up with a generic way of defining the functionality of a optical device. One approach is to describe the coupling between a set of input and output modes, since any linear optical device can be described in this fashion \cite{dabmiller_oe2012}.  This is particularly useful for waveguide-coupled devices, whose functionality can be defined in terms of the guided modes of the input and output waveguides.

In our design method, we specify device functionality by describing the mode conversion efficiency between a set of input modes and output modes. The input and output modes are specified by the user, and kept fixed during the optimization process. The input modes $i = 1 \ldots M$ are at frequencies $\omega_i$, and can be represented by an equivalent current density distribution $\mathbf{J}_i$. The fields $\mathbf{E}_i$ produced by each input mode satisfy Maxwell's equations,
\begin{align}
\nabla \times \mu_0^{-1} \nabla \times \mathbf{E}_i - \omega^2 \epsilon \mathbf{E}_i = - i \omega_i \mathbf{J}_i
\label{eqn:wgSplit3S_maxwell}
\end{align}
where $\epsilon$ is the permittivity distribution, and $\mu_0$ is the permeability of free space. 

For each input mode $i$, we then specify a set of output modes $j = 1 \ldots N_i$, whose amplitudes are bounded between $\alpha_{ij}$ and $\beta_{ij}$. If our output modes are guided modes of waveguides with modal electric fields $\mathcal{E}_{ij}$ and magnetic fields $\mathcal{H}_{ij}$, this constraint can be written using a mode orthogonality relationship \cite{prmcisaac_ieeetmtt1991},
\begin{align}
\alpha_{ij} \leq \left| \iint \left( \mathbf{E}_i \times \mathcal{H}_{ij} + \mathcal{E}_{ij} \times \mathbf{H}_i  \right) \cdot \hat{n} \, d\mathbf{r}_\perp \right| \leq \beta_{ij}.
\label{eqn:wgSplit3S_mode_constr}
\end{align}
Here, $\hat{n}$ is a unit vector pointing in the propagation direction, and $\mathbf{r}_\perp$ denotes the coordinates perpendicular to the propagation direction. We can use Faraday's law $\nabla \times \mathbf{E}_i = -i \omega \mu_0 \mathbf{H}_i$ to rewrite (\ref{eqn:wgSplit3S_mode_constr}) purely in terms of the electric field:
\begin{align}
\alpha_{ij} \leq \left| \iint \left( \mathbf{E}_i \times \mathcal{H}_{ij} + \mathcal{E}_{ij} \times \frac{i}{\omega \mu_0} \nabla \times \mathbf{E}_i  \right) \cdot \hat{n} \, d\mathbf{r}_\perp \right| \leq \beta_{ij}.
\label{eqn:wgSplit3S_mode_constrE}
\end{align}

More generally, we can specify the output mode amplitude in terms of a linear functional $\mathcal{L}_{ij}$ of the electric field $\mathbf{E}_i$, 
\begin{align}
\alpha_{ij} \leq \, \left| \mathcal{L}_{ij}\left(\mathbf{E}_i\right)\right| \, \leq \beta_{ij},
\label{eqn:wgSplit3S_L_constr}
\end{align}
where $V = \left\{\mathbf{E}:\mathbb{R}^3 \rightarrow \mathbb{C}^3 \right\}$ is the space of all possible electric field distributions, and $\mathcal{L}_{ij}:V \rightarrow \mathbb{C}$ maps electric field distributions to a complex scalar.

We are thus interested in finding a permittivity distribution $\epsilon$ and electric fields $\mathbf{E}_i$ which simultaneously satisfy (\ref{eqn:wgSplit3S_maxwell}) and (\ref{eqn:wgSplit3S_L_constr}), for all input modes $i = 1 \ldots M$ and output modes $j = 1 \ldots N$. To ensure that the resulting device is fabricable, we will later impose additional constraints on $\epsilon$.

\subsection{Linear algebra description}
Since we will solve Maxwell's equations numerically, and employ numerical optimization techniques to design our devices, it is convenient to recast the design problem in terms of linear algebra. We do this by discretizing space and making the substitutions
\begin{alignat}{2}
\mathbf{E}_i & \rightarrow x_i &&\in \mathbb{C}^n \nonumber \\
\epsilon & \rightarrow z &&\in \mathbb{C}^n \nonumber \\
\nabla \times \mu_0^{-1} \nabla \times & \rightarrow D && \in \mathbb{C}^{n \times n} \nonumber \\
 - i \omega_i \mathbf{J}_i & \rightarrow b_i && \in \mathbb{C}^{n} \nonumber \\
\mathcal{L}_{ij} & \rightarrow c_{ij} &&\in \mathbb{C}^n.
\end{alignat}
We thus wish to find electric fields $x_i$ and a permittivity distribution $z$ which satisfy
\begin{gather}
D x_i - \omega_i^2 \diag(z) x_i - b_i = 0 \label{eqn:wgSplit3S_maxwell_linalg}\\
\alpha_{ij} \leq \left| c^\dagger_{ij} x_i \right| \leq \beta_{ij} 
\end{gather}
for $i = 1 \ldots M$ and $j = 1 \ldots N_i$. Here, $\diag\left(v\right)$ refers to the diagonal matrix whose diagonal entries are given by the vector $v$, and $u^\dagger$ is the conjugate transpose of $u$. For convenience, we further define the matrices
\begin{align}
A_i(z)   &= D - \omega_i^2 \diag(z) \nonumber \\
B_i(x_i) &= - \omega_i^2 \diag(x_i) \label{eqn:wgSplit3S_defAB}.
\end{align}
This lets us rewrite equation (\ref{eqn:wgSplit3S_maxwell_linalg}) as
\begin{align}
0 = A_i(z) x_i - b_i = B_i(x_i) z + \left( D x_i - b_i \right).
\end{align}
The final problem we wish to solve is then
\begin{gather}
A_i(z) x_i - b_i = 0 \label{eqn:wgSplit3S_prob_maxwell}\\
\alpha_{ij} \leq \left| c^\dagger_{ij} x_i \right| \leq \beta_{ij} \label{eqn:wgSplit3S_prob_constr}
\end{gather}
for $i = 1 \ldots M$ and $j = 1 \ldots N_i$.

\subsection{Parametrizing the structure}
As described in the main article, we describe our structure using a two-dimensional level-set function $\phi(x,y):\mathbb{R}^2 \rightarrow \mathbb{R}$ , where the permittivity in the design region is given by
\begin{align}
\epsilon(x,y) = 
\begin{cases}
  \epsilon_1 & \mathrm{for} \; \phi(x,y) \leq 0 \\
  \epsilon_2 & \mathrm{for} \; \phi(x,y) > 0. \\
\end{cases}
\end{align}
For the purposes of numerical optimization, we discretize the level set function in space, which transforms the level set function into a two dimensional array $\phi \in \mathbb{R}^{U \times V}$. 

We parametrize the permittivity distribution $z$ with the level set $\phi$ by using a mapping function $m:\mathbb{R}^{U \times V} \rightarrow \mathbb{C}^n$, where
\begin{align}
z = m(\phi).
\end{align}
When the level set boundaries are not perfectly aligned with simulation grid cells, we render the structure using anti-aliasing. This allows us to continuously vary the structure, rather than being forced to make discrete pixel-by-pixel changes.

\subsection{Formulating the optimization problem}
We are finally in a position to construct our optimization problem. Although there are a variety of ways we could solve (\ref{eqn:wgSplit3S_prob_maxwell}) and (\ref{eqn:wgSplit3S_prob_constr}), the particular optimization problem we choose to solve is
\begin{alignat}{3}
&\minimize  \quad && F\left(x_1, \ldots , x_M\right)  \nonumber \\
&\subjectto \quad && A_i(z) x_i - b_i = 0, \quad \mathrm{for} \; i = 1 \ldots N_i  \nonumber \\
&                  && z = m(\phi). 
\label{eqn:wgSplit3S_optprob}
\end{alignat}
Here, we constrain the fields to satisfy Maxwell's equations, parameterize the permittivity $z$ with the level set function $\phi \in \mathbb{R}^{U \times V}$, and construct a penalty function
\begin{align}
F\left(x_1, \ldots , x_M\right) = \sum_{i = 1}^{M} f_i(x_i)
\label{eqn:wgSplit3S_fi}
\end{align}
for violating our field constraints from equation (\ref{eqn:wgSplit3S_prob_constr}). The penalty $f_i(x_i)$ for each input mode is given by
\begin{align}
f_i = \sum_{j = 1}^{N_i} I_+\left( \left|  c_{ij}^\dagger x_i \right| - \alpha_{ij} \right) + I_+ \left(\beta_{ij} - \left| c_{ij}^\dagger x_i \right| \right)
\end{align}
where $I_+\left(u\right)$ is a relaxed indicator function \cite{sboyd_2004},
\begin{align}
I_+\left(u\right) =
\begin{cases}
0, & u \geq 0 \\
\dfrac{1}{s} \left| u \right|^q, & \mathrm{otherwise}.
\end{cases}
\end{align}
Typically, we use $q = 2$ and $s = \max_i f_i(x_i) $.

\subsection{Steepest descent optimization}
We solve our optimization problem (\ref{eqn:wgSplit3S_optprob}) by first ensuring that Maxwell's equations  (\ref{eqn:wgSplit3S_prob_maxwell}) are always satisfied. This implies that both the fields $x_1, \ldots, x_M$ and the field-constraint penalty $F$ are a function of the level set $\phi$. We then optimize the structure using a steepest descent method. Using the chain rule, the gradient of the penalty function $F$ is given by
\begin{align}
\frac{dF}{d\phi} &= \frac{dF}{dz} \, \frac{dz}{d\phi} = \frac{dF}{dz} \, \frac{d}{d\phi} m(\phi)
\end{align}
since $z = m(\phi)$. The majority of the computational cost comes from computing the gradient $dF/dz$.

As described in the main text, we evolve the level set function $\phi$ by advecting it with a velocity field $v \in \mathbb{R}^{U \times V}$.  To implement gradient descent, we set the velocity field to be equal to the gradient of the penalty function:
\begin{align}
v = \frac{dF}{d\phi}.
\end{align}

\subsection{Computing gradient of penalty function $F$}
We now consider how to efficiently compute the gradient of the penalty function $F$ with respect to the permittivity $z$, which can be written using (\ref{eqn:wgSplit3S_fi}) as
\begin{align}
\frac{dF}{dz} = \sum^M_{i = 1} \frac{d}{dz} f_i (x_i).
\end{align}
Although $f_i$ is not a holomorphic function since $f_i: \mathbb{C}^n \rightarrow \mathbb{R}$, we can compute $d f_i / dz$ using the expression
\begin{align}
\frac{d}{dz} f_i (x_i) = 2 \operatorname{Re}\left( \frac{\partial f_i}{\partial x_i} \frac{dx_i}{dz} \right)
\label{eqn:wgSplit3S_dfidz_init}
\end{align}
where we have taken the Wirtinger derivative of $f_i$ \cite{lsorber_siamjo2012}. The Wirtinger derivative with respect to some complex variable $w = u + i v$ is defined as
\begin{align}
\frac{\partial}{\partial w} = \frac{1}{2} \left(\frac{\partial}{\partial u} - i \frac{\partial}{\partial v} \right) .
\end{align}
Using this definition, the Wirtinger derivatives $\partial f_i / \partial x_i$ are given by
\begin{align}
\frac{\partial f_i}{\partial x_i} = \sum_{j = 1}^{N_i} \frac{\partial}{\partial x_i} I_+ \left( \left| c_{ij}^\dagger x_i \right| - \alpha_{ij} \right) + \frac{\partial}{\partial x_i} I_+ \left( \beta_{ij} - \left| c_{ij}^\dagger x_i \right| \right)
\end{align}
where 
\begin{align}
&\frac{\partial}{\partial x_i} I_+ \left( \left| c_{ij}^\dagger x_i \right| - \alpha_{ij} \right)  =  \frac{1}{2} \frac{\left(c_{ij}^\dagger x_i\right)^\ast}{\left| c_{ij}^\dagger x_i \right|}\, c_{ij}^\dagger \; \nonumber \\
&\qquad\times
\begin{cases}
0, &  | c_{ij}^\dagger x_i | - \alpha_{ij} \geq 0 \\
\dfrac{q}{a} \left| | c_{ij}^\dagger x_i | - \alpha_{ij} \right|^{q - 1}, & \mathrm{otherwise}
\end{cases} \label{eqn:wgSplit3S_dIdx_1}\\
&\frac{\partial}{\partial x_i} I_+ \left( \beta_{ij} - \left| c_{ij}^\dagger x_i \right| \right)  =  \frac{1}{2} \frac{\left(c_{ij}^\dagger x_i\right)^\ast}{\left| c_{ij}^\dagger x_i \right|}\, c_{ij}^\dagger \; \nonumber \\
&\qquad\times
\begin{cases}
0, &  \beta_{ij} - | c_{ij}^\dagger x_i | \geq 0 \\
\dfrac{q}{a} \left| \beta_{ij} - | c_{ij}^\dagger x_i | \right|^{q - 1}, & \mathrm{otherwise}.
\end{cases} \label{eqn:wgSplit3S_dIdx_2}
\end{align}
Here, we have used the identity
\begin{align}
\frac{\partial}{\partial u} \left|u\right| = \frac{u^\ast}{2 \left| u \right| }.
\end{align}

Next, we consider how to take the derivative of the electric fields $x_i$ with respect to the permittivity $z$. If we take the derivative of the discretized Maxwell's equations (\ref{eqn:wgSplit3S_maxwell_linalg}) with respect to $z$, we obtain
\begin{align}
D \frac{dx_i}{dz} - \omega_i^2 \diag(x_i) - \omega_i^2 \diag(z) \frac{dx_i}{dz} &= 0  \nonumber \\
\left( D - \omega_i^2 \diag(z) \right) \frac{dx_i}{dz} &= \omega_i^2 \diag (x_i)  \nonumber \\
A_i(z) \frac{dx_i}{dz} &= - B_i(x_i)
\end{align}
where we have used our definitions of $A_i$ and $B_i$ from (\ref{eqn:wgSplit3S_defAB}). Rearranging, we find the derivative to be
\begin{align}
\frac{dx_i}{dz} &= - A^{-1}_i(z) B_i(x).
\label{eqn:dxi_dz}
\end{align}

We obtain our final expression for $df_i/dz$ by subsitututing (\ref{eqn:dxi_dz}) into (\ref{eqn:wgSplit3S_dfidz_init}):
\begin{align} 
\frac{d}{dz} f_i(x_i) &= 2 \operatorname{Re} \left( \frac{\partial f_i}{\partial x_i} \frac{d x_i}{dz} \right) \nonumber \\
&= 2 \operatorname{Re} \left( - \frac{\partial f_i}{\partial x_i} A_i^{-1}(z) B_i(x_i) \right) \nonumber \\
&= 2 \operatorname{Re} \left( - \left( A_i^{-\dagger}(z) {\frac{\partial f_i}{\partial x_i}}^\dagger  \right)^\dagger B_i(x_i) \right). \label{eqn:wgSplit3S_dfidz_final}
\end{align}
Since $A_i$ and $B_i$ are large $n \times n$ matrices, we have rearranged the expression in the final step to require only a single matrix solve rather than $n$ solves. This method for reducing the computational cost of computing gradients is known as \emph{adjoint sensitivity analysis}, and is described in detail elsewhere \cite{mbgiles_ftc2000}.

The cost of computing $dF/dz$ is dominated by the cost of solving Maxwell's equations. For each input mode $i = 1 \ldots M$, we need to solve both the \emph{forward} problem $x_i = A_i^{-1} b$ to find the electric field $x_i$, and the \emph{adjoint} problem $ A_i^{-\dagger}(z) {\frac{\partial f_i}{\partial x_i}}^\dagger $ from equation (\ref{eqn:wgSplit3S_dfidz_final}). Both the forward and adjoint problems can be solved by any standard Maxwell's equation solver \cite{nknikolova_itmtt,lalau-keraly_oe2013,aniederberger_oe2014}. We use a graphical processing unit (GPU) accelerated implementation of the finite-difference frequency-domain (FDFD) method \cite{wshin_jcp2012, wshin_oe2013}.

\newpage
\section{Level set implementation}
\subsection{Curvature limiting}
In the main text, we wrote that we implement curvature limiting by evolving the level set function $\phi$ with
\begin{align}
\phi_t - b(\kappa) \kappa \left| \nabla \phi \right| = 0
\label{eqn:wgSplit3S_curvfilt}
\end{align}
using the weighting function
\begin{align}
b(\kappa) = 
\begin{cases}
  1 & \mathrm{for} \; |\kappa| > \kappa_0\\
  0 & \mathrm{otherwise}. \\
\end{cases}
\end{align}
In practice, this has terrible convergence since the weighting function falls off infinitely sharply as the local curvature crosses $\kappa_0$. To improve the behaviour of our PDE, we actually use a smoothed weighting function
\begin{align}
b(x,y) = \exp\left(- \kappa_0^2 \; d^2(x,y)\right),
\end{align}
where $d(x,y)$ is the Euclidean distance to the nearest element in the set $\Omega$,
\begin{align}
d(x,y) = \inf_{(\hat{x},\hat{y}) \in \Omega} \lVert (x,y) - (\hat{x},\hat{y})\rVert.
\end{align}
We choose $\Omega = \{ (x,y) | \kappa(x,y) > \kappa_0 \}$ to be the set of points with a local curvature greater than our threshold $\kappa_0$. The distance function $d(x,y)$ can be efficiently computed using the Euclidean distance transform commonly included in image processing libraries.

\subsection{Numerical implementation}
In our design algorithm, we apply gradient descent using the partial differential equation
\begin{align}
\phi_t + v(x,y) \left| \nabla \phi \right| = 0
\label{eqn:wgSplit3S_normprop}
\end{align}
where $v(x,y)$ is the local velocity, and apply curvature limiting with equation \ref{eqn:wgSplit3S_curvfilt}. We spatially discretize equation \ref{eqn:wgSplit3S_normprop} using Godunov's scheme, and equation \ref{eqn:wgSplit3S_curvfilt} using central differencing, as is common practice \cite{sosher_2003}. We discretize in the time dimension using Euler's method.

To ensure that our level set equations remain well behaved, we regularly reinitialize $\phi$ to be a signed distance function \cite{sosher_2003}, where $\left| \nabla \phi \right| \approx 1$. Most reinitialization schemes, however, result in subtle shifts in the interface locations, which can cause optimization to fail. We use Russo and Smerka's reinitialization scheme to avoid these issues \cite{grusso_jcp2000}.

\begin{figure*}[h!tb]
\centering
\begin{minipage}{.5\textwidth}
  \centering
  \includegraphics[width=\linewidth]{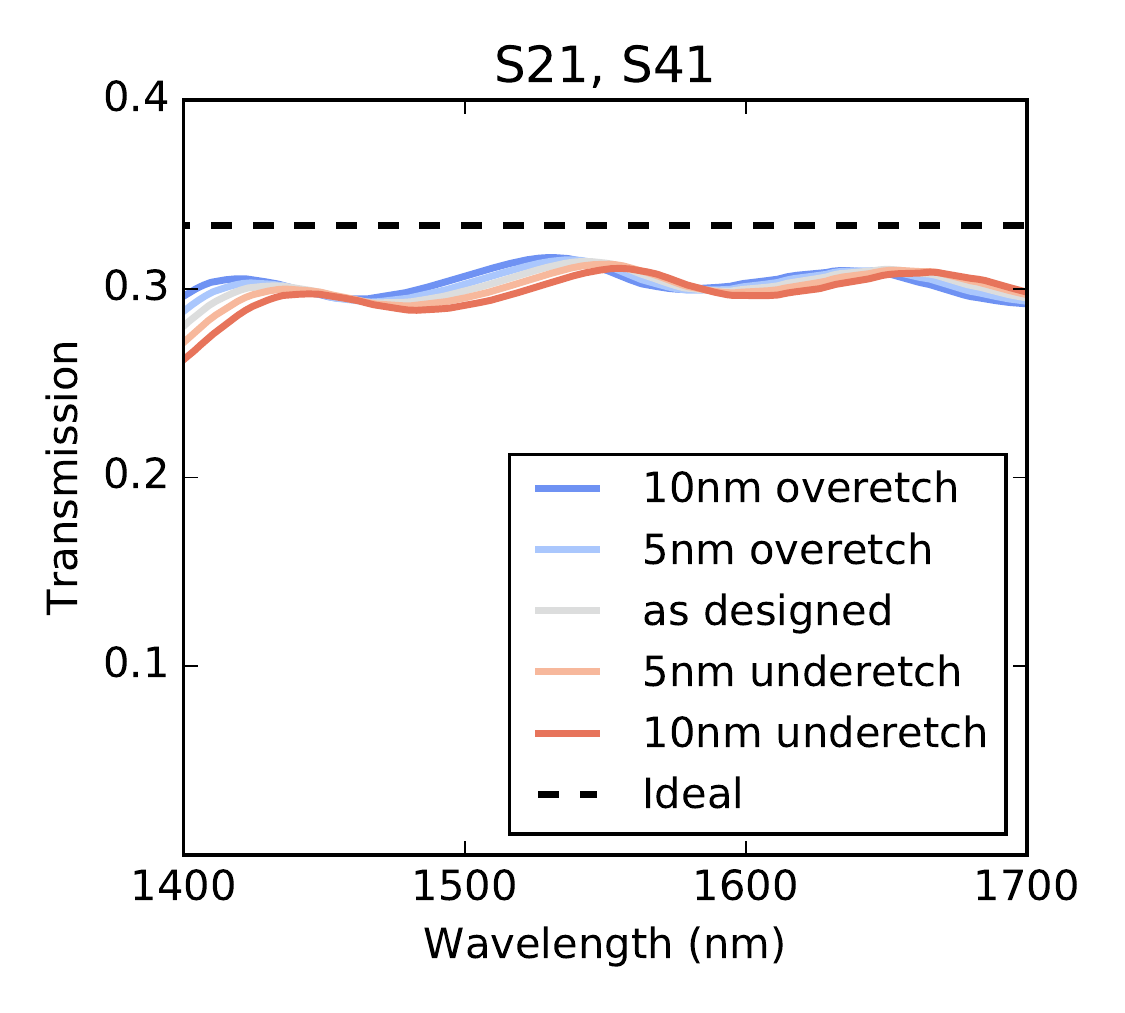}
\end{minipage}%
\begin{minipage}{.5\textwidth}
  \centering
  \includegraphics[width=\linewidth]{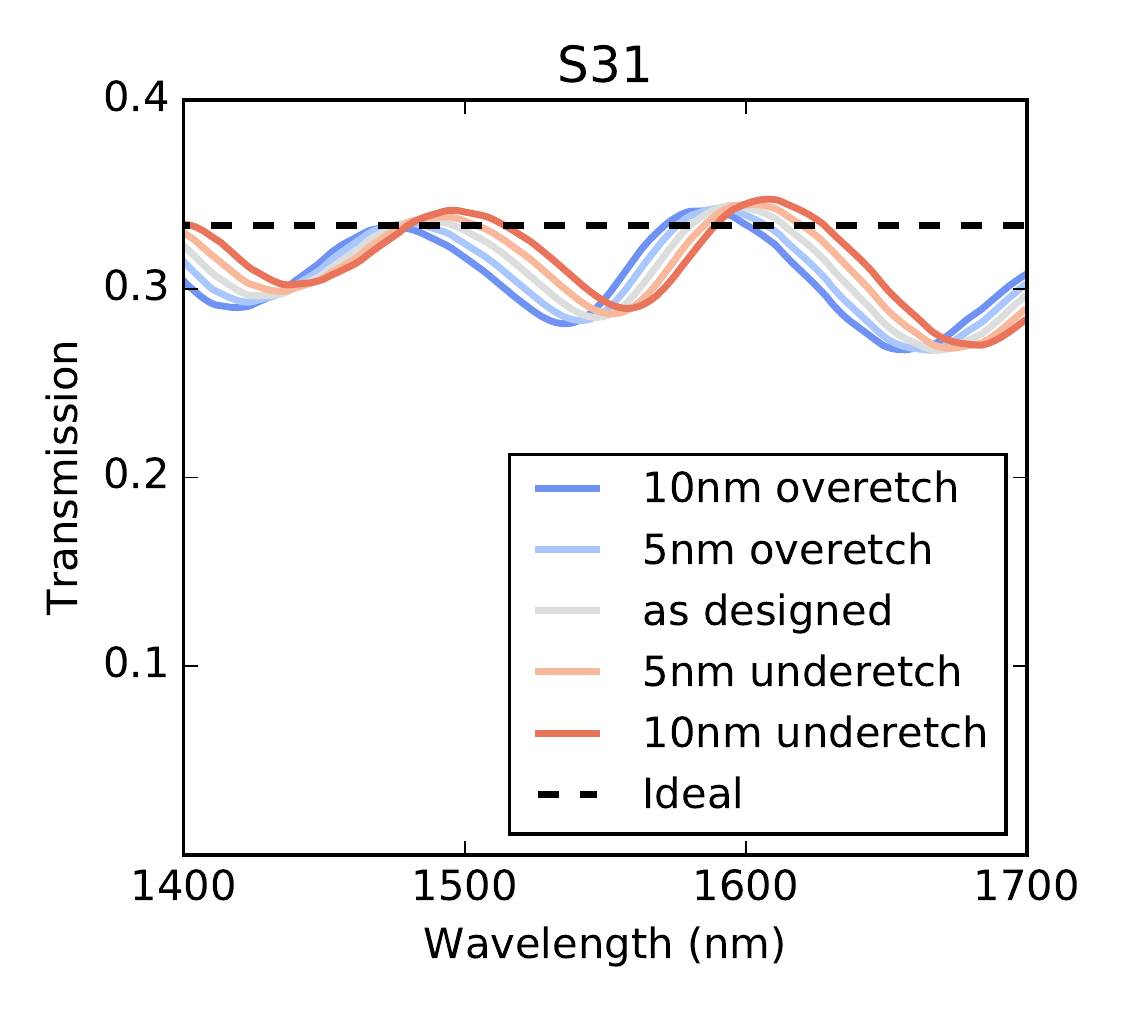}
\end{minipage}%
\caption{Simulated transmission spectra of the $1 \times 3$ splitter for a range of over-etching and under-etching errors, calculated using finite-difference time-domain (FDTD) simulations. For over-etching and under-etching of $10~\mathrm{nm}$, the device performance is not impacted except for slight spectral shifts. }
\label{fig:wgSplit3_tol}
\end{figure*}

\newpage
\section{Additional characterization of $1 \times 3$ splitter}
\subsection{Fabrication robustness}
To obtain a better understanding of the fabrication robustness of our $1 \times 3$ splitter, we simulated the device for a range of over-etching and under-etching errors, which correspond to lateral growth or shrinkage of the design. We have presented the results in figure \ref{fig:wgSplit3_tol}.

\subsection{Backreflections}
\begin{figure}[h!tb]
\centering
\includegraphics[width=\linewidth]{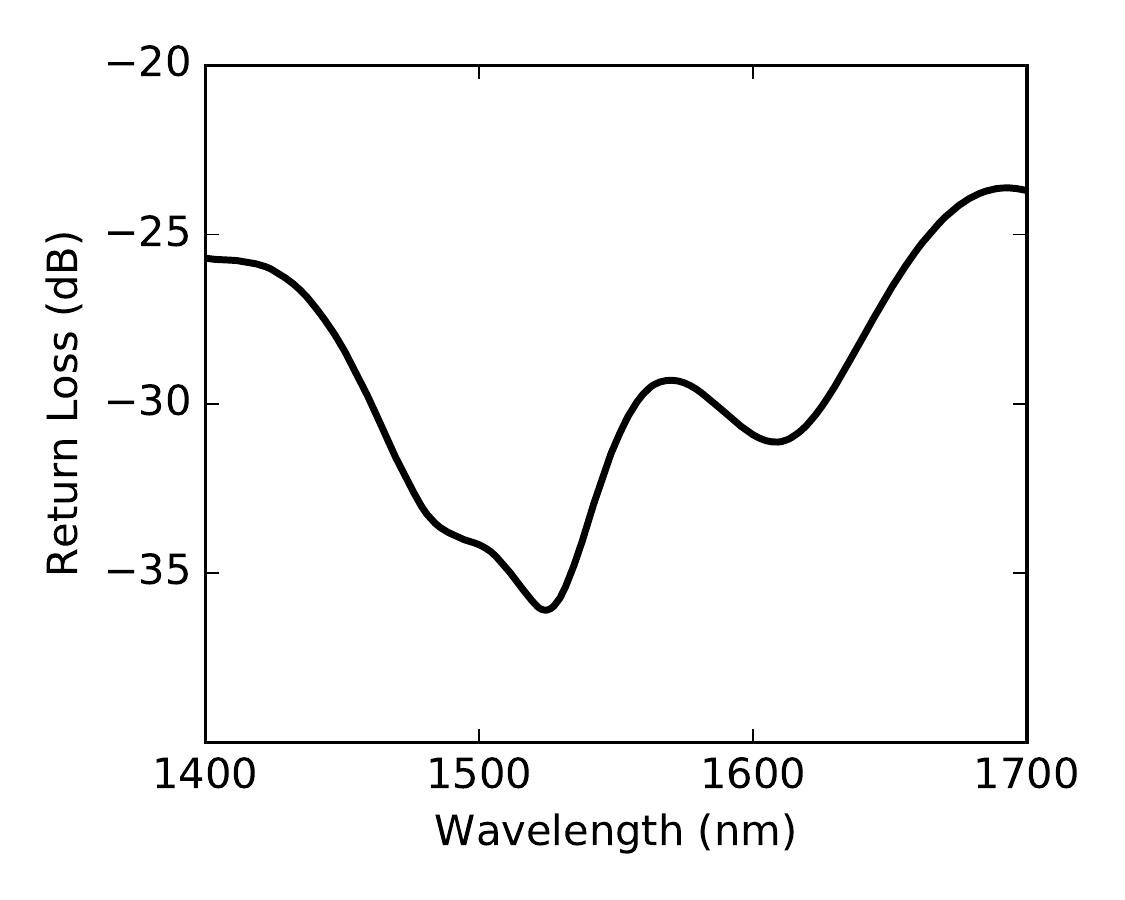}
\caption{Simulated return loss of the $1 \times 3$ splitter back into the fundamental mode of the input waveguide.}
\label{fig:wgSplit3_FDTD_RL}
\end{figure}

We also simulated the return loss for our $1 \times 3$ splitter, which we present in figure \ref{fig:wgSplit3_FDTD_RL}. The backreflections into the fundamental $\mathrm{TE}_{10}$ mode of the input waveguide are $< 23~\mathrm{dB}$ over the operating bandwidth of our device. Backreflections into the $\mathrm{TE}_{20}$ mode are zero due to reflection symmetry in our device in the horizontal direction, and mode conversion to TM modes is impossible due to reflection symmetry of our structure in the vertical direction. Finally, scattering into higher order waveguide modes is negligible since the input and output waveguides are close to single-mode. Thus, backreflections into the input waveguide comprise only a small fraction of the total losses.

\end{document}